# Effect of the quantum well thickness on the performance of InGaN photovoltaic cells

L. Redaelli[1,2,*], A. Mukhtarova[1,2], S. Valdueza-Felip[1,2], A. Ajay[1,2], C. Bougerol[1,3], C. Himwas[1,3], J. Faure-Vincent[1,4,5], C. Durand[1,2], J. Eymery[1,2], and E. Monroy[1,2]

[1] Université Grenoble Alpes, F-38000 Grenoble, France
[2] CEA-CNRS, Nanophysics and Semiconductors group, CEA-Grenoble, INAC-SP2M, F- 38000 Grenoble, France
[3] CEA-CNRS, Nanophysics and Semiconductors group, Institut Néel-CNRS, F-38000 Grenoble, France
[4] CNRS, INAC-SPRAM, F-38000 Grenoble, France
[5] CEA, INAC-SPRAM, F-38000 Grenoble, France
[*] luca.redaelli@cea.fr

**Abstract:** We report on the influence of the quantum well thickness on the effective band gap and conversion efficiency of $In_{0.12}Ga_{0.88}N$/GaN multiple quantum well solar cells. The band-to-band transition can be redshifted from 395 to 474 nm by increasing the well thickness from 1.3 to 5.4 nm, as demonstrated by cathodoluminescence measurements. However, the redshift of the absorption edge is much less pronounced in absorption: in thicker wells, transitions to higher energy levels dominate. Besides, partial strain relaxation in thicker wells leads to the formation of defects, hence degrading the overall solar cell performance.

InGaN alloys are considered as promising candidates for high-efficiency photovoltaic devices [1-4] since their band gap spans almost the whole solar spectrum from 0.7 eV (InN) to 3.4 eV (GaN). This makes theoretically possible the development of all-InGaN multijunction solar cells with a freely customizable number of junctions to enhance the overall efficiency. However, the large lattice mismatch between GaN and InN has led several groups to study the possibility of hybrid integration, combining an InGaN cell in a tandem device with silicon [5,6] or other non-III-nitride [7] photovoltaic cells.

The difficulty of growing high-quality InGaN layers increases with the In content. Reports of InGaN-based junctions with an In mole fraction exceeding 0.3 are rare [1]; the best external quantum efficiencies (EQEs) exceeding 0.7 are obtained at around 400 nm and quickly drop for longer wavelengths [8-10]. The main challenges are the large dislocation density and In-clustering, caused by the strong tendency to phase separation during growth. Absorbing layers in the form of a multiple quantum well (MQW) structure are often used to delay strain relaxation. Furthermore, the quantum confined Stark effect (QCSE) associated to the strong piezoelectric fields in the InGaN/GaN system [11] offers the possibility to tune the effective band gap of the structure by adjusting the quantum well (QW) and barrier thickness ($t_{QW}$ and $t_B$, respectively). The effect of tuning $t_B$ in InGaN/GaN MQW photovoltaic devices has been studied by Wierer *et al.* [12] and Watanabe *et al.* [13]. According to their results, the absorption cutoff of the solar cells redshifts with decreasing $t_B$. However, this does not always translate in enhanced overall cell efficiency, since the short circuit current density ($J_{sc}$) and open circuit voltage ($V_{oc}$) also depend on $t_B$.

In this paper, we focus on the influence of the QW thickness on the effective band gap of the junction and its impact on the overall cell efficiency. We experimentally demonstrate that the band-to-band transition in InGaN QWs can be significantly redshifted in larger QWs. However, this redshift appears linked to a dramatic enlargement of the Stokes shift, so that increasing the $t_{QW}$ above a few nm is no longer beneficial. This fact, combined to the increased defect density associated to strain relaxation in thicker QWs, causes thick-QW solar cells to have inferior overall performance.

The samples considered in this study consist of a non-intentionally-doped $In_{0.12}Ga_{0.88}N$ /GaN MQW ($t_B$ = 8.5 nm, $t_{QW}$ = 1.3- 5.4 nm) active region sandwiched between p- and n-type GaN contact layers. Figures 1(a-c) show the band diagram of three QWs in the middle of the active region calculated at zero bias using the nextnano[3] software [14,15]. The QWs present the characteristic sawtooth profile due to the polarization-induced internal electric field. The band-to-band transition between the topmost hole energy level ($h_1$) and the deepest electron level ($e_1$) decreases in energy with increasing $t_{QW}$, hence the expected absorption wavelength redshifts from 390 nm to 475 nm, as shown by the solid line in Fig. 1(d). Furthermore, when $t_{QW}$ increases, the overlap of the hole and electron wavefunctions decreases, and the deepening of the potential wells enhances the carrier localization. It is important to note that the calculations in Figs. 1(a-c) were performed under the assumption of fully-strained wells. Growing thicker QWs increases the probability of lattice relaxation: as the energy shift is mostly due to the piezoelectric component of the polarization, relaxation would result in a smaller band gap redshift. The calculated redshift for full relaxation is shown by the dashed line in Fig. 1(d).

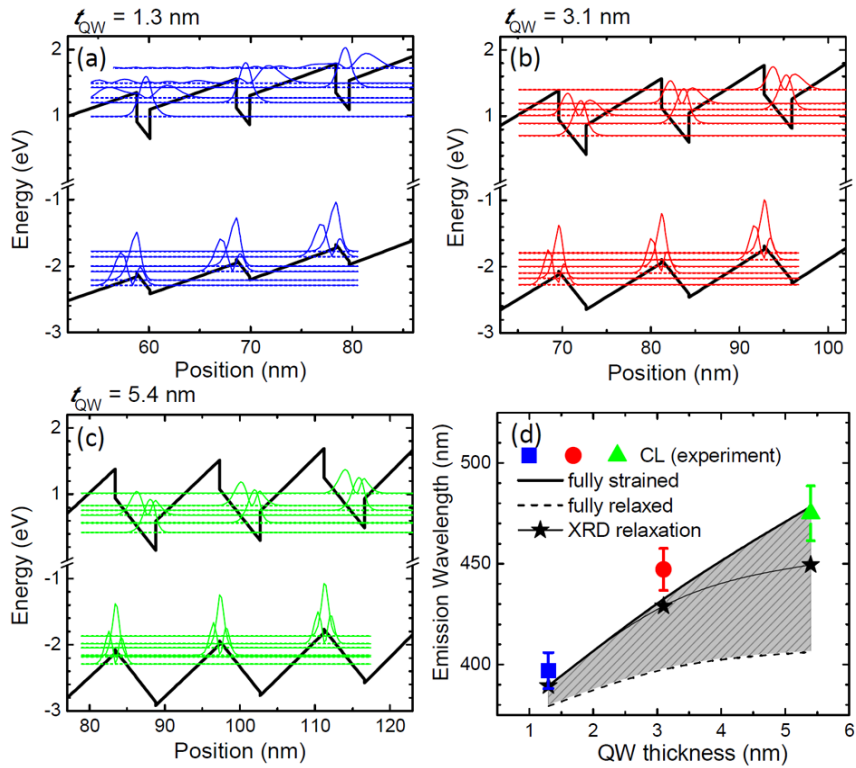

Fig. 1: Band diagrams of the $In_{0.12}Ga_{0.88}N$/GaN ($t_B$ = 8.5 nm) MQW structure for (a) $t_{QW}$ = 1.3 nm, (b) $t_{QW}$ = 3.1 nm, (c) $t_{QW}$ = 5.4 nm. The two first electronic levels in each QW are shown with their squared wavefunctions. These calculations assume the QWs fully strained on GaN, and a lattice temperature of 5 K. The n-doped layers are towards the left side, the p-layers towards the right. (d) Calculated $e_1$-$h_1$ transition wavelength as a function of the QW thickness assuming the MQW fully strained on GaN and fully relaxed. The stars indicate the result for the strain state estimated by XRD measurements. The calculations are compared to the CL peak emission wavelength measured at 5 K. The error bars indicate the full widths at half maximum of the CL peaks.

In order to experimentally investigate these effects, three samples with different QW thickness were grown on *c*-plane sapphire by metal-organic vapor phase epitaxy. The active region consists of a 30-period $In_{0.12}Ga_{0.88}N/GaN$ MQW with $t_B$ = 8.5 nm and $t_{QW}$ = 1.3, 3.1, and 5.4 nm, as in the previously presented simulations. The GaN:Mg top layer has a thickness of 80–110 nm and a hole concentration in the $10^{17}$ $cm^{-3}$ range. In the following we will refer to the three samples as A ($t_{QW}$ = 1.3 nm), B ($t_{QW}$ = 3.1 nm) and C ($t_{QW}$ = 5.4 nm), as summarized in Table I. Further details about the growth of the samples are reported in Ref. [16].

High-resolution transmission electron microscopy (HR-TEM) experiments were performed on a *Jeol 3010* microscope operated at 300 kV. The strain state and In content of the samples were measured by x-ray diffraction (XRD): θ-2θ scans around the (0004) Bragg peak reflection were recorded at the BM02 beamline of the European Synchrotron Radiation Facility using a wavelength λ = 1.28361 Å, and reciprocal space maps around the (-1015) reflection were measured in an XRD diffractometer using the λ =1.54056 Å Cu K$_α$ line. Cathodoluminescence (CL) measurements were performed in a *FEI quanta 200 CL* system, using an acceleration voltage of 10 kV and a current of ~100 pA. The sample emission was collected by a parabolic mirror and focused onto a *Jobin Yvon HR460* monochromator equipped with a charge-coupled device and a photomultiplier tube. The absorption spectra were deduced from transmission measurements at normal incidence, neglecting reflection. Excitation was provided by a 150 W Xe-arc lamp coupled to a *Gemini-180* double monochromator for the ultraviolet spectral range (250–400 nm), and by a 1000 W halogen lamp coupled with an *Omni Lambda 300* monochromator for the visible range (360–700 nm), both calibrated with a reference Si photodetector.

Samples were processed in photovoltaic devices with mesa sizes of 1×1 $mm^2$ and 0.5×0.5 $mm^2$ defined by $Cl_2$-based inductively coupled plasma etching. The n-contact surrounding the mesas consisted of e-beam evaporated Ti/Al/Ni/Au. The p-contact consisted of a semitransparent Ni/Au (5 nm/5 nm) layer, annealed for 5 min in oxygen-rich atmosphere, and of a 130 nm-thick Ni/Au finger structure (finger width 5 μm, pitch 150 μm). Further details about the fabrication process have been reported in Ref. [16]. The spectral response of the cells was measured at room temperature and normal incidence in the same setups used for transmission measurements. A He-Cd laser was used to determine the EQE at 325 nm. Solar cell measurements were performed under 1 sun illumination (AM1.5G irradiation with total power of 1000W/m²) using a class AAA solar simulator from Oriel.

Cross-sectional HR-TEM micrographs of the samples' MQW are shown in Fig. 2. In sample A [Fig. 2(a)], the periodicity of the MQW is very uniform and the layer interfaces are sharp. In sample B the MQW is also mostly smooth and uniform [Fig. 2(b)], although a few areas present evidence of elastic and plastic misfit relaxation. This degradation is clearly observed on much larger areas of sample C, as illustrated in Figs. 2(c) and (d). The values of $t_B$ and $t_{QW}$ measured from the HR-TEM images are summarized in Table I.

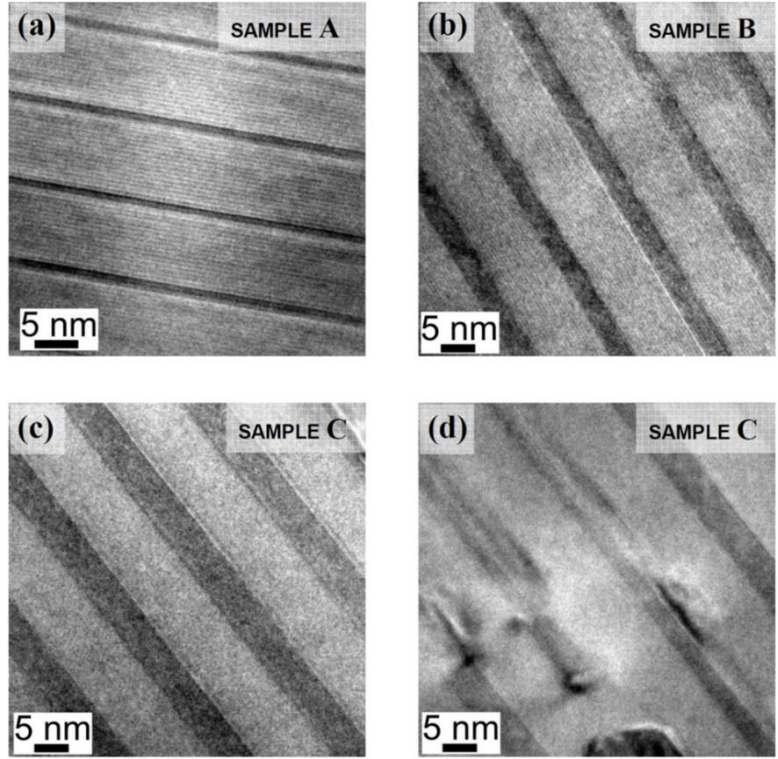

Fig. 2: HR-TEM micrographs of the MQW region of samples (a) A, (b) B, and (c and d) C. In (c), a section of sample C with smooth wells and sharp interfaces is shown, while in (d) the MQW is disturbed due to strain relaxation

Table I: Quantum well and barrier thickness, and MQW period thickness for the samples under study, determined from the HR-TEM micrographs and XRD investigations, respectively. Photovoltaic performance of the fabricated solar cells.

| Sample | HR-TEM $t_B$ (nm) | $t_{QW}$ (nm) | HR-XRD MQW period (nm) | Cell parameters $J_{sc}$ (mA/cm$^2$) | $V_{oc}$ (V) | FF (%) | $\eta$ (%) |
|---|---|---|---|---|---|---|---|
| A | 8.0 ± 0.26 | 1.3 ± 0.26 | 9.7 | -0.42 ± 0.008 | 2.17 ± 0.05 | 54 ± 2 | 0.49 ± 0.014 |
| B | 9.3 ± 0.26 | 3.1 ± 0.26 | 12.0 | -0.46 ± 0.045 | 1.29 ± 0.08 | 42 ± 2 | 0.25 ± 0.043 |
| C | 9.1 ± 0.26 | 5.4 ± 0.26 | 13.7 | -0.14 ± 0.002 | 1.14 ± 0.01 | 47 ± 3 | 0.07 ± 0.002 |

In order to assess the structural quality of the MQW on a larger scale, the samples were investigated by XRD. In Fig. 3(a), the θ-2θ scans show several satellites of the MQW reflection, indicating well-defined multilayer structures in spite of the local degradation observed by HR-TEM in samples B and C. The average MQW periods determined from the distance between satellites are summarized in Table I. The In content and the degree of relaxation of the MQW were estimated from reciprocal space maps around the (-1015) reflection. For samples A and B (not shown here), the reflections corresponding to the GaN layers and to the MQW are vertically aligned (same in-plane wave vector $Q_x$), which means that the MQWs are fully strained on GaN within the experimental error bars [17]. In the case of sample C (Fig. 3(b)), combining the information from the (-1015) and (0004) reflections with the layer thicknesses measured by HR-TEM, a degree of relaxation of 40±15% is estimated. The average In content in the QWs is $x$ = 0.12±0.02 for the three samples.

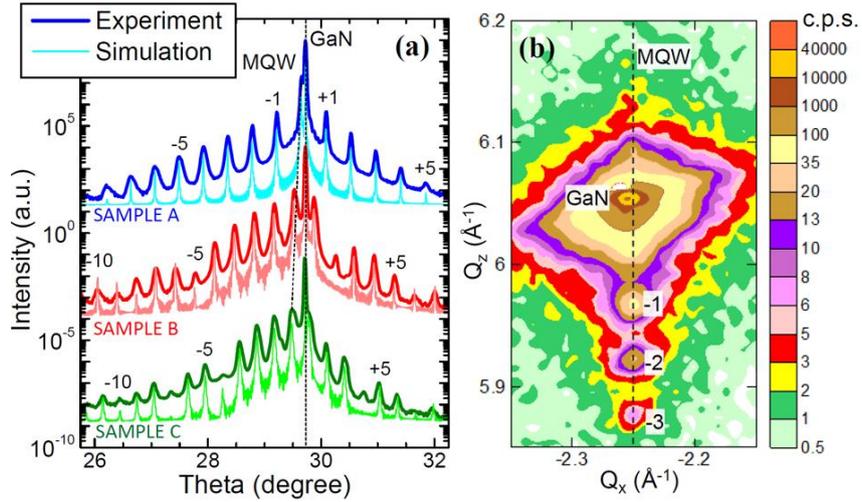

*Fig. 3: (a) XRD ϑ-2ϑ scans around the (0004) reflection of samples A, B and C compared to simulations performed by the X'Pert Epitaxy software from Panalytical assuming 12% In content and the layer thickness reported in Table I. (b) Reciprocal space map around the (-1015) reflection of sample C (c.p.s. = counts per second).*

Figure 4 displays low temperature (5 K) CL measurements of samples A and C. The CL spectra in the insets of Fig. 4 show a distinct emission line which redshifts with increasing QW thickness. The peak values for the three samples are plotted in Fig. 1(d), showing good agreement with the simulated trend. The experimental data appear slightly redshifted in comparison with the calculated values taking partial relaxation into account, which could be explained by the error bars in the determination of the In composition and by the presence of alloy and thickness fluctuations in the QWs. Such inhomogeneities are confirmed by the nonuniform emission in large QWs observed in monochromatic CL maps (see comparison between samples A and C in Fig. 4).

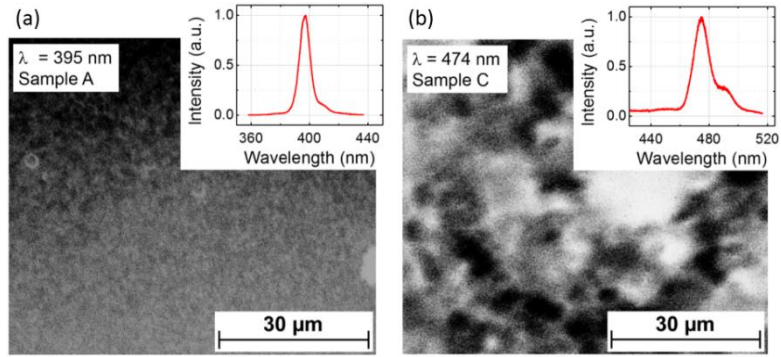

*Fig. 4: Monochromatic low-temperature (5K) CL maps of samples A and C at the peak emission wavelength. In the insets, the respective spectra are shown.*

Figure 5(a) shows the spectral response of the solar cells under study. For each sample, at least three solar cells taken from different areas of the wafer were measured; the error bars associated to the responsivity measurement at 325 nm are ±1% for sample A, ±12% for sample B and ±2.5% for sample C. No systematic difference was observed in the performance of 1×1 mm$^2$ and 0.5×0.5 mm$^2$ solar cells. Considering the best devices, peak responsivities at 370 nm of 155 mA/W (EQE = 0.52), 107 mA/W (0.36), and 39 mA/W (0.13) were obtained for samples A, B and C, respectively. The efficiency drop is attributed to increased nonradiative recombination, caused by the larger defect density observed by HR-TEM in thicker-QW devices. Fig. 5(b) shows the variation of the spectral cutoff wavelength as a function of $t_{QW}$, compared with theoretical calculations of the transitions between the first electron and hole levels ($e_1$-$h_1$) and the second electron and hole levels ($e_2$-$h_2$). Sample A fits well the calculations for $e_1$-$h_1$. As expected, the absorption cutoff of sample B is redshifted with respect to A, although not as much as theoretically predicted. For the thickest QWs (sample C) no further shift of the absorption edge is observed with respect to B, in contrast with the large spectral shift observed in CL measurements. However, its absorption cutoff corresponds well with the value calculated for the $e_2$-$h_2$ transition.

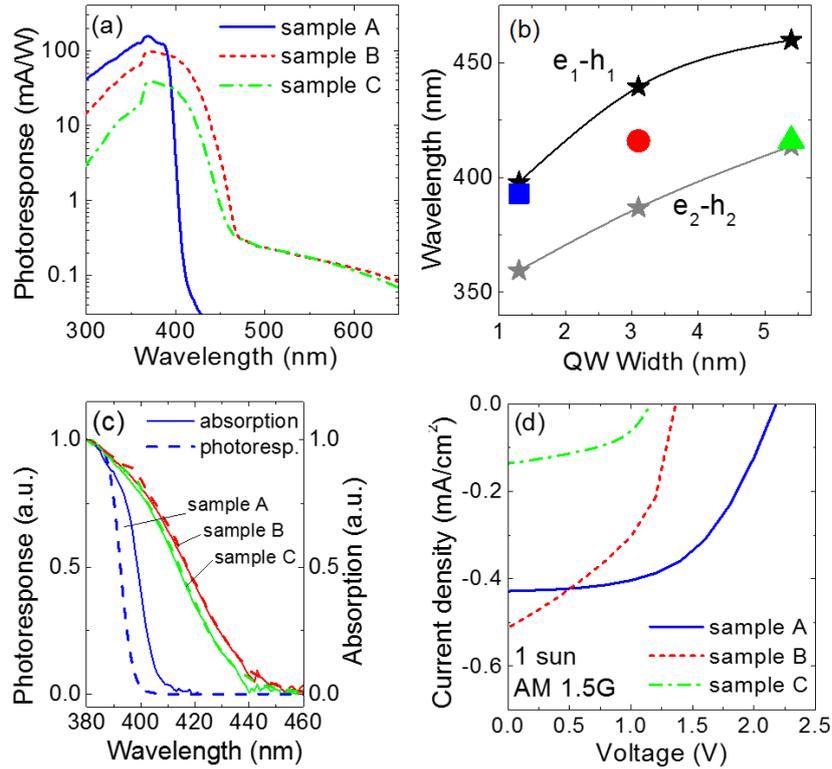

*Fig. 5: (a) Spectral photoresponse of the devices under study. The curves correspond to the best performing 0.5×0.5 mm2 solar cell of the respective sample. (b) Calculated e1-h1 and e2-h2 transition wavelengths as a function of the QW width compared to the experimental spectral cutoff of the samples (defined as the wavelength where the photocurrent drops to half its maximum value). The calculations are performed with nextnano3 assuming the lattice at room temperature and the strain states measured by XRD. (c) Comparison between photoresponse (dashed lines) and absorption (full lines) spectra, normalized at 382 nm. (d) J–V characteristics under 1 sun AM1.5G equivalent illumination [same solar cells as in (a)].*

To understand the cause of the missing redshift, in Fig. 5(c) the spectral response of the three samples is compared to their spectral absorption. All curves are normalized at 382 nm, in order to compare only the InGaN MQW absorption and exclude the contribution of the GaN layers. The good overlap of the curves relating to samples B and C indicates that the collection losses are negligible, since the absorption and photoresponse cutoffs coincide. The missing absorption redshift can hence be explained by the low oscillator strength of the $e_1$-$h_1$ transition, which decreases drastically with the QW thickness due to the QCSE, combined with the lower density of states of the ground levels compared to excited levels, e.g. $e_2$, $h_2$. In CL measurements the emission is dominated by the ground state transition e1-h1, due to the fast intraband relaxation of excited carriers. However, when increasing the QW thickness, the contribution of $h_1$-$e_1$ to the absorption becomes negligible in comparison to the contribution of excited levels [18] e.g. $h_2$-$e_2$, with higher density of states and better electron-hole wavefunction overlap. This trend is visible in Fig. 1(a) to (c), and causes the $h_2$-$e_2$ transition to dominate the absorption in sample C.

Figure 5(d) depicts the current density-voltage (*J-V*) curves of the solar cells under 1 sun irradiance. Sample B has the largest $J_{sc}$, which can be explained by the linear increase of the sun's irradiance from 300 nm to 450 nm and the redshift of the absorption cutoff. In sample B, the better overlap of the photoresponse with the solar spectrum largely compensates the drop in peak responsivity with respect to sample A. However, $V_{oc}$ and the shunt resistance (inverse of the tangent to the *J-V* curve in *V* = 0 V) are lower in sample B, resulting in a lower fill factor (*FF*). As a consequence, the maximum efficiency (*η*) of the cells fabricated on sample B is only 0.25±0.04%, which is about half the value obtained on sample A (0.49±0.01%). The loss in $V_{oc}$ in sample B points to a leakage path through the active region and/or to increased nonradiative recombination, which might be associated to the increased defect density. $J_{sc}$ decreases for sample C because of the much lower external quantum efficiency with respect to sample B. An overview of the main cell parameters is provided by Table I.

In summary, it has been shown that the band-to-band transition in In$_x$Ga$_{1-x}$N QWs with an In mole fraction *x* = 0.12 can be redshifted more than 70 nm by increasing the QW thickness from 1.3 nm to 5.4 nm. However, this redshift cannot be fully exploited to improve the efficiency of solar cells, due to the reduced overlap of the ground state electron and hole wavefunctions caused by the QCSE in thicker QWs. Besides, thicker wells are prone to the generation of defects in the MQW structure because of strain and local misfit relaxation, which results in lower peak EQE, together with reduced $V_{oc}$ and shunt resistance.

**Acknowledgments:** Authors would like to acknowledge financial support by the Marie Curie IEF grant "SolarIn" (#331745) and by the French National Research Agency via the GANEX program (ANR-11-LABX-0014). We acknowledge the ESRF for provision of synchrotron radiation facilities.